\documentclass[aps, pra, twocolumn, superscriptaddress, floatfix]{revtex4-1}
\usepackage{amsmath}
\usepackage{amsfonts}
\usepackage{braket}
\usepackage[hidelinks]{hyperref}
\usepackage{graphicx}

\newcommand{\UMBC}{Department of Physics, University of Maryland Baltimore County, Baltimore, Maryland 21250, USA}

\begin{document}

\title{Local Gradient Optimization of Leakage-Suppressing Entangling Sequences}

\author{A.~A.~Setser}
\affiliation{\UMBC}
\author{J.~P.~Kestner}
\affiliation{\UMBC}

\begin{abstract}
We use a gradient-based optimization scheme to find single-qubit rotations to be interwoven between timesteps of a noisy logical two-qubit entangling gate in order to suppress arbitrary logical and leakage errors in the two-qubit gate. We show how the sequence fidelity is affected by imperfections in the single-qubit operations, as well as by various relative strengths of the logical and leakage noise. Our approach is completely general and system-independent, allowing for application to any two-qubit system regardless of the experimental implementation details.
\end{abstract}

\maketitle

\section{Introduction}\label{intro}
Reliable implementation of two-qubit entangling gates is a key step towards creating a useful quantum computer. In order for fault-tolerant quantum computing to be possible, operations on qubits must be performed with error rates less than the ``error correction threshold", with the exact value of the threshold varying with the qubit encoding scheme. One of the highest error correction thresholds is around $1\%$, offered by surface codes \cite{WangPRA2011}. However, it is still desirable to suppress errors as much as possible, in order to reduce the surface code overhead.

The difficulty in physical implementation arises when environmental effects are taken into consideration. Interactions between the system and environment can entangle the two, causing a collapse of the wavefunction, thus destroying the quantum properties of the system. These are called incoherent errors, and cannot be reversed via unitary operations on the system. However, system-environment interactions can also introduce random perturbative effects within the system Hamiltonian, causing the coherent evolution of the qubit to differ from the unperturbed evolution. Such effects are known as coherent errors, and it is possible to dynamically correct these errors via unitary operations on the system \cite{CumminsPRA2003, JonesNMR2011}.  In this work, we focus on such suppression of coherent errors.

For single qubits, coherent errors are often suppressed through a wide variety of composite pulse sequences, which typically consist of piecewise constant values of the system parameters chosen such that the errors incurred during each timestep cancel with those of the other timesteps in the evolution, causing the final evolution to be error-free up to a given order in the perturbation \cite{yangDD2010, merrillBook, vandersypenRMP2005}.
For two-qubit entangling gates, that approach is complicated by the larger dimensionality of the Hilbert space. However, the potential benefit is even greater than in the single-qubit case because the errors are generally larger.  That is due to the longer evolution times required for two-qubit entangling gates since the qubit-qubit interaction term in the system Hamiltonian is often weak compared to the single-qubit terms.

It is known that high-fidelity single-qubit operations can be used as a resource in two-qubit pulse sequences to suppress arbitrary coherent errors within the two-qubit logical subspace. These two-qubit pulse sequences are typically found analytically and are restricted to either small numbers of single-qubit rotations, or allow for larger numbers of single-qubit operations at the cost of restricting them to simple $\pi$ rotations for simplicity \cite{Tomita2010, cohenPRA2016, JonesPRA2003, ichikawaPRA2013, FernandoPRL2017, hillPRL2007}. However, this ease of experimental implementation typically comes at the cost of reduced error-suppression. We have demonstrated this fact previously, where we showed that a numerical optimization method can be used to find the optimal single-qubit rotations to suppress the arbitrary logical errors \cite{armanPRA}. It was shown via simulations that the numerically optimized sequences were generally more effective at suppressing arbitrary logical errors than the analytically derived sequences, at the cost of being somewhat more difficult to implement experimentally since the single-qubit rotations were not restricted to simple rational multiples of $\pi$.

While our previous method focused on suppressing logical errors only, we are also interested in the possibility of suppressing leakage errors, as may be the case in a system like a superconducting qubit with multiple energy levels \cite{krantzAPR2019, wendinRPP2017, petererPRL2015}. 
In this paper, we present a variant of the numerical optimization scheme developed in Ref.~\cite{armanPRA} that now addresses both leakage and logical errors simultaneously.  We optimize single-qubit rotations inserted between applications of a noisy two-qubit entangling gate such that the final sequence performs a logical entangling operation while suppressing all coherent errors. Despite the fact that the inserted rotations are restricted to act within the logical subspace whereas the error acts in a much larger space, our method is surprisingly effective. We also show that it is relatively unaffected by imperfections in the interwoven single-qubit operations and we demonstrate how the performance changes with varying logical and leakage noise strengths. The modular, system-independent nature of our approach facilitates application to any two-qubit setup, regardless of the details of the Hamiltonian.

\section{Model and Optimization Scheme}\label{model}
We model each qubit as a two level system, coupled to a third leakage level of higher energy. This could be, for example, an excited state of a weakly anharmonic superconducting qubit or an excited valley state of a silicon spin qubit. Although more than one leakage level may exist, the population of such levels becomes increasingly unlikely as the energy of the leakage level increases. We therefore consider only a single leakage level. In order for the numerical optimization scheme to be effective for a wide range of noise values, we sample $M$ total noise realizations and require that the optimized single-qubit operations perform well over the average of these realizations \cite{GoerzPRA2014}. 

Within a noise realization $m$, we assume that there exists an evolution operator $U^{(m)}$, such that the system state $\ket{\psi(t)}$ evolves according to $\ket{\psi(T)}=U^{(m)}\ket{\psi(0)}$, where $T$ is the gate duration. The single-qubit evolution operators reside in SU(3), which is generated by the Gell-Mann matrices, $\lambda_i$, with $i\in[0, 8]$. The forms of the matrices are shown in Appendix \ref{gellmann}. We denote the upper $2\times2$ block of $U^{(m)}$ as the logical subspace. The logical subset of operations is generated by $\lambda_1, \lambda_2, \text{ and } \lambda_3$, since they are respectively equal to the Pauli matrices $\sigma_X, \sigma_Y, \text{ and } \sigma_Z$ within the logical subspace and are zero outside of it. Leakage effects are generated by $\lambda_4$ through $\lambda_8$, since they contain terms which couple the logical subspace to the leakage subspace. 

Following Ref.~\cite{armanPRA}, we choose a time evolution operator composed of a series of $N$ time steps with the form
\begin{equation}\label{U}
    U^{(m)}=\prod_{n=N}^1 \exp\left[-\frac{i\pi}{N}\lambda_{3, 3}\right]\exp\left[-\frac{i}{N}\Delta^{(m)}\right]R_n,
\end{equation}
where $\lambda_{i, j}=\lambda_i\otimes\lambda_j$ and the $R_n$'s are arbitrary single-qubit rotations. Within each step of the evolution, $\exp\left[-i\pi\lambda_{3, 3}/N\right]$ is equal to the Nth root of a $2\pi$ conditional phase gate within the logical subspace and an identity operation outside of it. Noise is introduced through the term $\exp\left[-i\Delta^{(m)}/N\right]$, where 
\begin{equation}\label{delta}
    \Delta^{(m)}=\sum_{ij}\delta^{(m)}_{i,j}\lambda_{i, j}
\end{equation}
and $\delta_{i,j}$ is a random error coefficient which acts on the error channel $\lambda_{i,j}$. The factor of $1/N$ in the exponential of the error term reflects the fact that we expect the error to scale with the size of the time slice.

The form for the evolution operator essentially consists of splitting a noisy $2\pi$ conditional phase gate into $N$ timesteps and inserting arbitrary single-qubit rotations in between them. We denote such a sequence of operations as a length-$N$ sequence. The single-qubit rotations steer the evolution dynamics, so that the final interaction between qubits in the logical subspace is not limited to the $\sigma_{ZZ}$ interaction that would be generated in the absence of the single-qubit rotations. We choose a Pauli vector parametrization for the single-qubit operations,
\begin{align}
    R_n&=\exp\left[i\left(\alpha_{1,n}\lambda_1+\beta_{1.n}\lambda_2+\gamma_{1,n}\lambda_3\right)\right] \notag \\
    &\otimes \exp\left[i\left(\alpha_{2,n}\lambda_1+\beta_{2.n}\lambda_2+\gamma_{2,n}\lambda_3\right)\right], \label{R}
\end{align}
where $\alpha_{1,n}\ldots \gamma_{2,n}$ are the free parameters for optimization.

The optimal free parameters are the ones that minimize the optimization function we choose. There are two requirements for our functional: reduce noise in the final operation (both logical and leakage) and generate a perfect entangler within the logical subspace. A perfect entangler is a gate which can produce a maximally entangled state from an unentangled one \cite{PhysRevA.82.034301}. The suppression of error can be measured by first finding the fidelity of the gate $U^{(m)}$ and taking the target gate to be the noise-free version of $U^{(m)}$, i.e., 
\begin{equation}\label{F}
    F\left(U^{(m)}\right)=\frac{1}{81}\left|\text{tr}\left(O^{\dagger}U^{(m)}\right)\right|^2,
\end{equation}
where
\begin{equation}\label{O}
    O=\prod_{n=1}^{N}\exp\left[-\frac{i\pi}{N}\lambda_{3,3}\right]R_n. 
\end{equation}
The gate error is then 
\begin{equation}\label{gate error}
    \varepsilon\left(U^{(m)}\right)=1-F\left(U^{(m)}\right).
\end{equation}
Note that the target gate includes the single-qubit rotations and is therefore changing over the course of the optimization. Optimization of the gate error with respect to the free parameters will produce a known final operation which is robust against noise.

Although this would produce an error-free operation, it is unlikely to be a perfect entangler within the logical subspace, since the single-qubit rotations will affect the entanglement dynamics. Since $U^{(m)}$ is in SU(9), we quantify the qubit-qubit entanglement it produces by first projecting onto the SU(4) logical subspace in order to obtain a nonunitary effective logical evolution, then examining the Makhlin invariants of the projected operation, $g_1, g_2, \text{ and } g_3$ \cite{Makhlin2002}.  Although this definition is strictly valid only when the total leakage vanishes, since the leakage is being minimized via Eqs. \ref{F} and \ref{O}, it is an effective way to quantify entangling power within the cost function for the purposes of optimization. Specifically, the distance between a logical two-qubit operation and the nearest perfect entangler can be expressed according to the Makhlin invariants as \cite{WattsPRA2015}
\begin{equation}\label{d}
    d=g_3\sqrt{g_1^2+g_2^2}-g_1.
\end{equation}

This distance measure can take on negative values for certain operations, which can be problematic for the optimization routine \cite{armanPRA}. To check when these problematic operations occur, we calculate the quantity
\begin{equation}\label{s}
    s=\pi-\cos^{-1}(z_1)-\cos^{-1}(z_3)
\end{equation}
from the ordered roots $(z_1, z_2, z_3)$ of the equation \cite{ZhangPRA2003}
\begin{equation}
    z^3-g_3z^2+\left(4\sqrt{g_1^2+g_2^2}-1\right)z+\left(g_3-4g_1\right)=0,
\end{equation}
and note that the actual distance metric that should be minimized in order to realize a logical perfect entangler is 
\begin{equation}\label{D}
    \mathcal{D}\left(U^{(m)}\right) =
    \begin{cases}
  d  & \text{$d>0$ and $s>0$} \\
  -d & \text{$d<0$ and $s<0$} \\
  0  & \text{otherwise.}
  \end{cases}
\end{equation}
This is a true metric, in the sense that it is positive for non-perfect entanglers and equal to zero for perfect entanglers.

The total functional for optimization is the sum of the gate error and the distance to the nearest perfect entangler averaged over all noise realizations,
\begin{equation}\label{functional}
J=\frac{1}{M}\sum_{m=1}^M\varepsilon\left(U^{(m)}\right)+\mathcal{D}\left(U^{(m)}\right).
\end{equation}
Minimization of the functional serves to produce a perfect entangler within the logical subspace that is robust against logical and leakage noise. 

We use the L-BFGS-B gradient-based minimization algorithm \cite{SIAMJSciComput.16.1190}, which is implemented within the SciPy optimization package \cite{scipy}. We choose a gradient-based minimization algorithm, since gradient-free algorithms are generally slower for larger numbers of optimization parameters, like we have in our scheme \cite{EPJQT.2.21}. The L-BFGS-B algorithm also offers an increase in convergence speed through the estimation of the Hessian of the functional. SciPy's implementation of the algorithm also allows the gradient of the functional to be estimated numerically, so we do not need to calculate the analytic gradient of $J$.

Since the L-BFGS-B algorithm is a \textit{local} search method, the convergence of the routine is highly dependent on the initial ``guess" parameters we choose at the start of the minimization. In order to choose an effective initial guess, if the greatest divisor of $N$ is $d$, we repeat the solution for the length $d$ sequence $N/d$ times to use as the guess for the length $N$ sequence. This ensures that longer-length sequences will be constructed from shorter-length sequences which have already been optimized according to our criteria. For prime length sequences, we set the guess for the free parameters to be all zeroes, so that the single-qubit operations are initialized as identity operations.

\section{Results}\label{results}
When evaluating the success of the optimization routine, we separately track our two criteria: generating a noise-free operation and generating a logical perfect entangler. The final gate error is evaluated via Eq. \ref{gate error}. The normalized fidelity of the gate $U^{(m)}$ with respect to the nearest perfect entangler is given by \cite{WattsPRA2015}
\begin{equation} \label{Fpe}
  F_{\text{PE}}(U^{(m)})=
  \begin{cases}
  \cos^2\left(\frac{c_1+c_2-\frac{\pi}{2}}{4}\right) & c_1+c_2 \leq \frac{\pi}{2} \\
  \cos^2\left(\frac{c_2+c_3-\frac{\pi}{2}}{4}\right) & c_2+c_3 \geq \frac{\pi}{2} \\
  \cos^2\left(\frac{c_1-c_2-\frac{\pi}{2}}{4}\right) & c_1-c_2 \geq \frac{\pi}{2} \\
  1 & \text{otherwise,}
  \end{cases}
\end{equation}
where $c_1, c_2, \text{ and } c_3$ are the Weyl chamber coordinates for $U^{(m)}$ \cite{ZhangPRA2003}. The error associated with the distance to the nearest perfect entangler, averaged over noise realizations, is then
\begin{equation}\label{pe error}
    \varepsilon_{\text{PE}}=\frac{1}{M}\sum_{m=1}^M 1-F_{\text{PE}}\left(U^{(m)}\right).
\end{equation}
We take the projection of our final optimized SU(9) operations onto the logical subspace in order to obtain an effective logical $4\times4$ evolution operator that can be used in Eq. \ref{pe error}. Since this quantity is between $0$ and $1$, it is a more direct measure of the final entanglement capabilities of the operations, compared to the metric $\mathcal{D}$ that was used in the optimization.

The error coefficients $\delta^{(m)}_{i,j}$ are drawn randomly from a normal distribution with a standard deviation of $\sigma_{\text{nonlocal}}=0.065$. This value is chosen so that when no rotations are inserted to suppress error, the gate error is around $10\%$, which is a realistic situation \cite{NPJQuantumInformation.3.3}. We find that $M=100$ is enough to ensure that our results are robust against a general noise realization, i.e., the optimized solutions obtained by running the routine over different sets of $100$ noise realizations do not change significantly.

\begin{figure}[t]
  \includegraphics[width=\columnwidth]{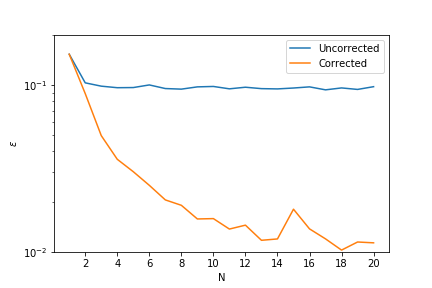}
  \caption{Gate error in relation to sequence length for a logical $\sigma_{ZZ}$ interaction, assuming access to perfect single-qubit rotations.}
  \label{gate error no local noise}
\end{figure}

The results of the optimization are shown in Figure \ref{gate error no local noise}. The gate error initially decreases rapidly with increasing $N$, but the gains diminish as the sequence grows longer. We do not know what causes this saturation at large $N$, but it is useful that order-of-magnitude improvements in gate error can be obtained already with $N\sim 10$.  While not shown, all sequences with $N>2$ have $\varepsilon_{\text{PE}}=0$. 
A supplementary data file is available which contains the solutions for all sequence lengths for this case and all further cases we consider \cite{armansRepository}.

We also consider the effects of imperfections in the local operations on the performance of the optimization routine. Like the two-qubit operations, we assume that the local operations have both logical and leakage noise. Noise within the logical subspace is modeled by introducing perturbations into the control parameters $\eta_i\in\{\alpha_{1,i}, \ldots, \gamma_{2,i}\}$ according to
\begin{equation}\label{perturb}
    \eta_i\rightarrow\eta_i^{\prime}=\eta_i(1+\delta_{\eta}),
\end{equation}
where $\delta_{\eta}$ is an error coefficient drawn randomly from a normal distribution with a standard deviation of $\sigma_{\text{local}}$. 

The leakage noise is introduced by multiplying the local operations $R_n$ by the factor
\begin{align}
    \prod_{k=4}^8 \exp\left(i\sqrt{\alpha_{1,n}^2+\beta_{1,n}^2+\gamma_{1,n}^2}\delta_{k}\lambda_k\right)\notag \\
    \otimes\exp\left(i\sqrt{\alpha_{2,n}^2 + \beta_{2,n}^2+\gamma_{2,n}^2}\delta_{k}^{\prime}\lambda_k\right),
\end{align}
where the $\delta_k$ and $\delta_k^{\prime}$ are error coefficients drawn randomly from a normal distribution, also taken to have a standard deviation of $\sigma_{\text{local}}$ for simplicity. The choice for the form of the leakage noise ensures that the errors introduced are proportional to the magnitude of the logical rotations being performed. This is a realistic situation, since larger rotations generally correspond to longer gate times and thus introduce more error. The standard deviation for the distribution of the local error coefficients is taken to be $\sigma_{\text{local}}=0.002$, so that the local rotations have a fidelity of approximately $99.9\%$ when calculated according to
\begin{equation} \label{F_R}
  F_R =
    \frac{1}{81} \left|
      \text{tr}\left(
        R^{\dagger}\left(
          \alpha^{\prime}_{1},\ldots,\gamma^{\prime}_{2}\right)
        R\left(\alpha_{1},\ldots,\gamma_{2}\right)
      \right)\right|^2\,,
\end{equation}
and averaged over $1000$ sets of error coefficients and $1000$ sets of angles drawn randomly from a uniform distribution ranging from $-2\pi$ to $2\pi$.

\begin{figure}[t]
  \includegraphics[width=\columnwidth]{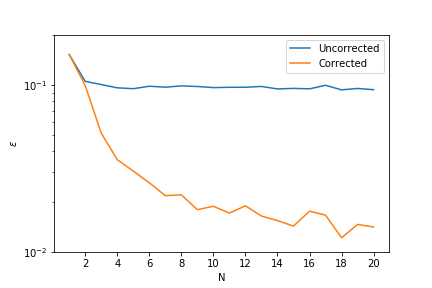}
  \caption{Gate error in relation to sequence length for a logical $\sigma_{ZZ}$ interaction, for the case of noisy single-qubit rotations}
  \label{gate error with local noise}
\end{figure}

The results of this optimization are shown in Figure \ref{gate error with local noise}. The scaling in this case is similar although slightly worse than the error-free local rotation case, achieving a maximum fidelity of $98.8\%$ compared to $99.0\%$. Again, though not shown here, all sequences with $N>2$ have $\varepsilon_{\text{PE}}=0$. Thus, imperfections in the local operations have only marginal effects at the small sequence lengths we consider. As the sequence length increases, previous work has shown that single-qubit errors can continuously increase to the point where gate errors begin to increase within increasing $N$ \cite{armanPRA}. Furthermore, the solutions obtained in the presence of local noise will perform just as well if the noise is nonlocal only, whereas solutions obtained in the absence of local noise will not significantly decrease the gate error if local noise is introduced.

\begin{figure}[t]
  \includegraphics[width=\columnwidth]{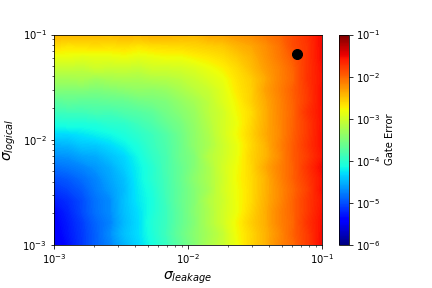}
  \caption{Gate error of the $N=16$ solution obtained in the absence of local noise, in relation to varying standard deviations for the logical and leakage noise. The solution obtained was optimized at $\sigma_{\text{logical}}=\sigma_{\text{leakage}}=0.065$ (marked on plot).}
  \label{contour plot}
\end{figure}
%comment: change axis labels to sigmas
%resolved

While the optimization producing Fig.~\ref{gate error no local noise} was performed assuming equal strength errors for the logical and leakage noise (i.e., that the $\delta_{i,j}$s of Eq. \ref{delta} are all drawn from the same distribution), we wish to see how the solutions hold as we separately vary these error strengths. Figure \ref{contour plot} shows the results of the $N=16$ solution obtained in the absence of local noise, in relation to varying standard deviations for the logical and leakage noise. The solution is taken from the Figure \ref{gate error no local noise} case, which was optimized with $\sigma_{\text{logical}}=\sigma_{\text{leakage}}=0.065$. From Figure \ref{contour plot}, we see that the optimized solution is more sensitive to leakage noise than logical noise. This is reasonable, since the terms in Eq. \ref{delta} which generate logical errors have $i,j\in\{0, 1, 2, 3\}$ (not including the identity term, $i=j=0$), while the rest of the terms generate leakage errors. This gives $15$ logical error generators and $65$ leakage error generators, making the overall sequence more susceptible to leakage errors.

\begin{figure}[t]
  \includegraphics[width=\columnwidth]{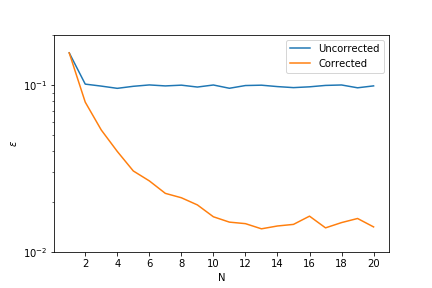}
  \caption{Gate error in relation to sequence length for a logical $\sigma_{XX}+\sigma_{YY}$ interaction, for the case of noisy single-qubit rotations with error-free logical $\sigma_Z$ rotations}
  \label{xx+yy plot}
\end{figure}

So far we have only considered a two-qubit $\lambda_{3,3}$ interaction in Eq. \ref{U}, i.e, a $\sigma_{ZZ}$ interaction in the logical subspace. However, the performance of the optimization routine is not limited strictly to this form for the interaction. We can also consider a logical $\sigma_{XX}+\sigma_{YY}$ interaction, which is relevant for many superconducting qubit setups \cite{ibmPRAPP2016}. This is reflected by changing $\lambda_{3,3}\rightarrow\lambda_{1,1}+\lambda_{2,2}$ in Eq. \ref{U}. In these types of systems, one can also typically perform ``virtual $Z$ gates," in which local logical $\sigma_Z$ rotations can be performed instantaneously in software by changing the reference phase of the microwave pulses that drive single-qubit rotations \cite{ibmPRA2017}. Allowing these error-free logical $\sigma_Z$ rotations in our optimization corresponds to setting $\delta_{\gamma_{1}}=\delta_{\gamma_{2}}=0$. (We have already observed that the optimization is not strongly affected by small local errors, so accounting for virtual gating actually doesn't make a big difference, but we do so just to show that it is not difficult to incorporate such considerations.) We again take $\sigma_{\text{nonlocal}}=0.065$ and $\sigma_{\text{local}}=0.002$.
The results of this optimization are shown in Figure \ref{xx+yy plot}. The performance of the optimization routine is similar to the logical $\sigma_{ZZ}$ interaction case, achieving a minimum gate error of $98.6\%$. As with the previous cases, all sequences with $N>2$ have $\varepsilon_{\text{PE}}=0$. Thus, the optimization routine is not significantly affected by changing the logical two-qubit interaction from $\sigma_{ZZ}$ to $\sigma_{XX}+\sigma_{YY}$.

\section{Conclusion}
We have shown that two-qubit logical entangling gates with fidelities around $90\%$ can be used in conjunction with logical single-qubit operations to construct two-qubit entangling gates with errors of around $1\%$. The logical single-qubit operations are interwoven between timesteps of the entangling operation and effectively suppress both arbitrary logical and leakage coherent errors present in the entangling gates.

We have shown that our numerical optimization is effective even when imperfections in the single-qubit operations are considered. In addition, we have shown that our method is effective both for a logical two-qubit $\sigma_{ZZ}$ interaction and a $\sigma_{XX}+\sigma_{YY}$ interaction. For the $\sigma_{ZZ}$ interaction case, we have shown how the optimized solutions depend individually on the strengths of the logical and leakage noise present. The modular nature of this approach allows for application to any two-qubit system, regardless of the Hamiltonian.

\section*{Acknowledgments}
This research was sponsored by the Army Research Office (ARO), and was accomplished under Grant Number W911NF-17-1-0287.

\appendix

\section{Gell-Mann Matrices}\label{gellmann}

For completeness, the Gell-Mann matrices are presented here. They are given by
\begin{align}
    &\lambda_0=
    \begin{pmatrix}
    1&0&0\\0&1&0\\0&0&1
    \end{pmatrix}, 
    \lambda_1=
    \begin{pmatrix}
    0&1&0\\1&0&0\\0&0&0
    \end{pmatrix}, 
    \lambda_2=
    \begin{pmatrix}
    0&-i&0\\i&0&0\\0&0&0
    \end{pmatrix},\notag \\
    &\lambda_3=
    \begin{pmatrix}
    1&0&0\\0&-1&0\\0&0&0
    \end{pmatrix},
    \lambda_4=
    \begin{pmatrix}
    0&0&1\\0&0&0\\1&0&0
    \end{pmatrix},
    \lambda_5=
    \begin{pmatrix}
    0&0&-i\\0&0&0\\i&0&0
    \end{pmatrix},\notag \\
    &\lambda_6=
    \begin{pmatrix}
    0&0&0\\0&0&1\\0&1&0
    \end{pmatrix},
    \lambda_7=
    \begin{pmatrix}
    0&0&0\\0&0&-i\\0&i&0
    \end{pmatrix},
    \lambda_8=
    \frac{1}{\sqrt{3}}\begin{pmatrix}
    1&0&0\\0&1&0\\0&0&-2
    \end{pmatrix}.\notag
\end{align}

\bibliography{bibfile.bib}

\end{document}